\documentstyle[twocolumn, epsbox]{article}
\begin{document}
\begin{center}
{\bf Long periodicity of Blasers with RXTE ASM}

{S.Osone, M.Teshima and K.Mase}

{Institute of Cosmic Ray Research, Universty of Tokyo, Kashiwa 277-8582, Japan}
\end{center}

%\correspondence{S.Osone(osone@icrr.u-tokyo.ac.jp)}

%\firstpage{1}
%\pubyear{2001}

% \titleheight{11cm} % uncomment and adjust in case your title block
                     % does not fit into the default and minimum 7.5 cm

%\maketitle

\begin{abstract}
We have searched for long periodicity in ten X-ray selected Blasers with {\it RXTE} ASM 4.6 years data. We found about 10-100 day possible periodicities for three TeV gamma ray Blasers, Mkn 421, Mkn 501 and PKS 2155-304.
\end{abstract}

\section{Introduction}
Begelman  et al.(1980) suggested that multiple Blackholes may exist at the center of AGN from the existence of merging galaxies. If there is a multiple Blackhole system, we may see an orbital period by a secondary Blackhole crossing the accretion disk of the primary or a precession of a jet. For example, a 12 years periodicity is confirmed in optical data with 100 years for a radio selected Blaser OJ 287(Sillanpaa et al. 1988).
Also, a periodicity of 23 days for X-ray selected Blaser Mkn 501 was detected with the TeV gamma ray cherenkov detector Telescope array (Hayashida et al. 1996) and HEGRA (eg. Aharonian et al. 1999a) and  with the Xray detector {\it RXTE} All Sky Monitor (ASM;Levine et al. 1996) during a TeV flare in 1997 (Kranich et al. 1999; Nishikawa et al.1999).

There is an observation of radio jet bending in Xray selected Blasers Mkn 421 (Piner et al.), Mkn 501 (Giovannini et al. 2000) with {\it VSOP}, which may suggest precession or orbital motion of the jet. There also may be a long periodicity from a thermal instability in an accretion disk (Honma et al. 1991; Abramowicz et al. 1995). If we scale the periodicity of 10 sec of Blackhole  binary GRS 1915$+$105 (eg. Morgan et al. 1997) to AGN of a mass with 10$^7$ $M_{\odot}$, a periodicity of 100 day is expected.

Here, we concentrate on Blasers which may have a periodicity from a geometrical origin. We aim to confirm a periodicity of 23 days for Mkn 501 and search for longer periodicities than 1 day for Blasers using the {\it RXTE} ASM archive data with a Lomb method.

\section{Observation}
We used {\it RXTE} ASM data from January 1996 (MJD 50087) to August 2000 (MJD 51780), (4.6 years).
We collected Xray selected Blasers cataloged by Xie et al.(1993) from  ASM archives, as Table 1. We obtained  a series of 90 sec integrated data. We show the lightcurve of three interesting Blasers which show possible periodicities in our analysis, in figure 1.

\begin{table}[h]
\caption{Target list of Xray selected Blasers }

\vspace{10pt}
\begin{tabular}{ccc} \hline
Mkn 501 & PKS 0548-322 & Mkn 421\\
  Mkn 180 & H 1101-232 &  BL 1ZW187\\
 BL 1426+427 & BL 1219+305 &  BL 0323+022\\
 PKS 2155-304 &  &  \\ \hline
\end{tabular}
\end{table}

\begin{figure*}[htb]
%\vspace{-60pt}
 \psbox[height=5cm,width=5cm]{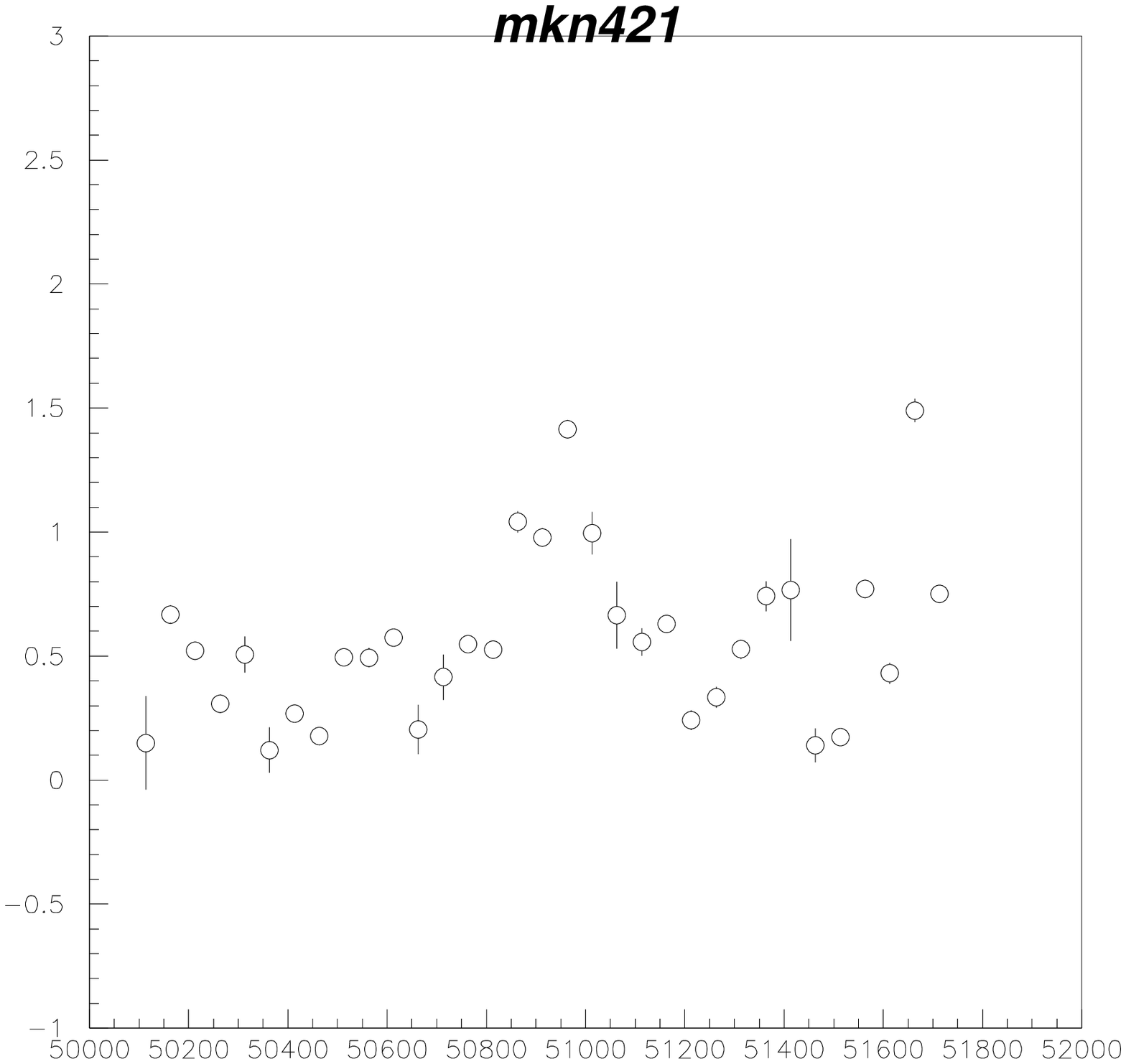}
%\hspace{-35pt}
 \psbox[height=5cm,width=5cm]{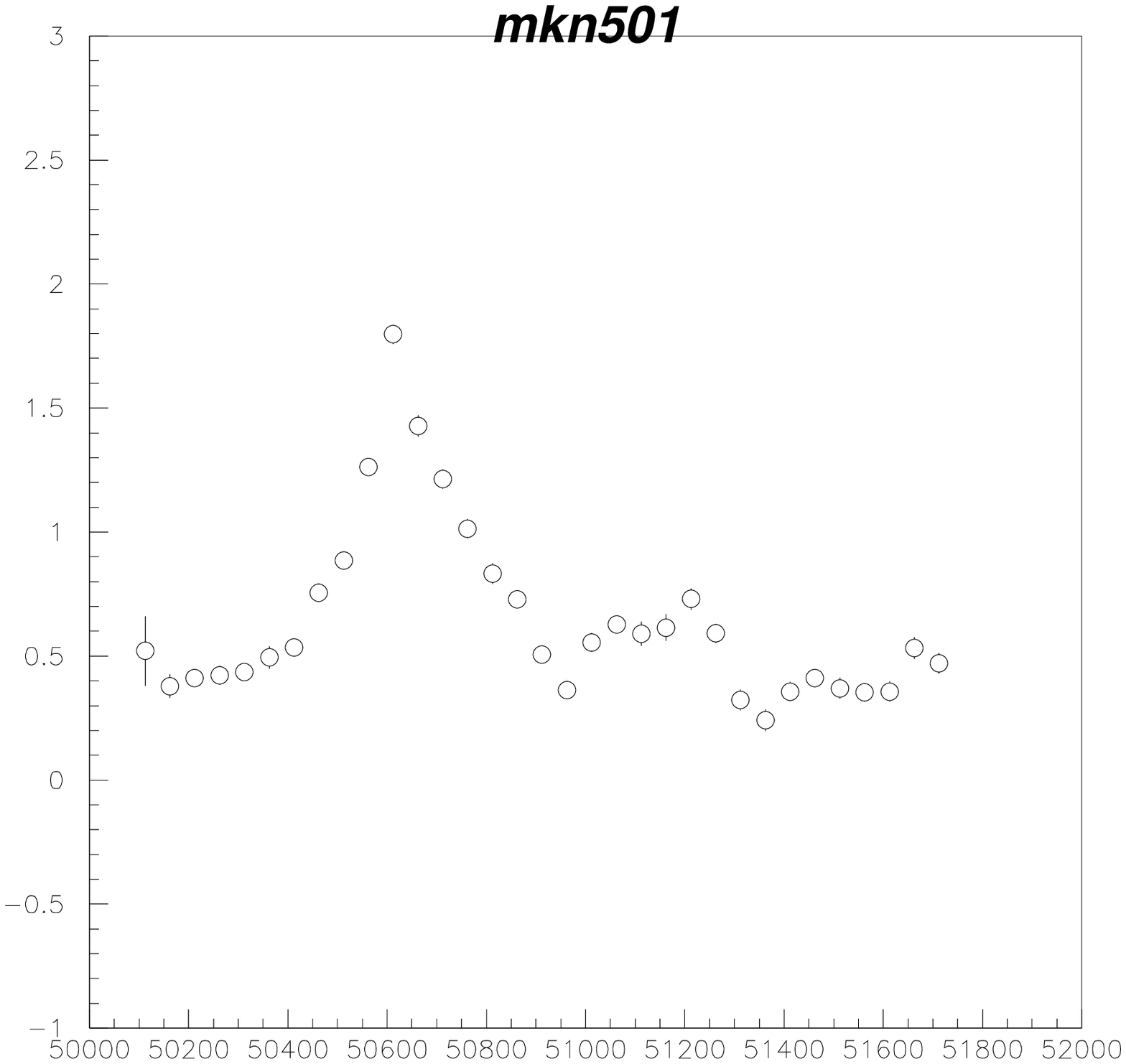}
%\hspace{-35pt}
 \psbox[height=5cm,width=5cm]{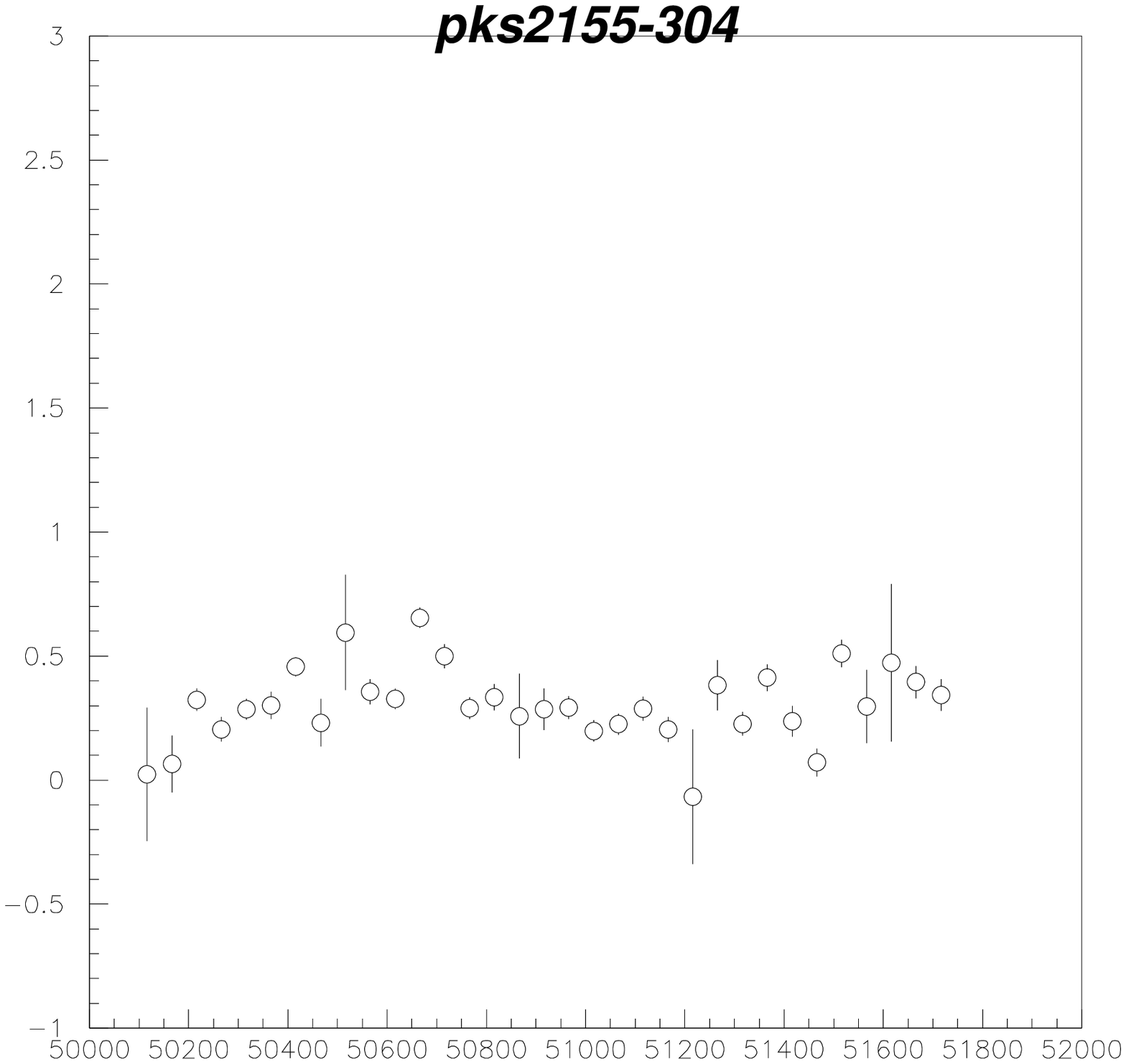}
 \caption{Lightcurve of three interesting Blasers in our analysis. A possible periodicity with chance probability less than 10$^{-3}$ is detected. Vertical axis is the count rate for each detector of ASM , horizon axis is the observation time in unit of MJD. Data is binned in 50 day increments and selected with data points $\ge$20 for clarity.}
\end{figure*}

\section{Analysis}

In order to examine the periodicity of X-ray intensities from each source, we use a Lomb method (Lomb 1976; Scrargle 1982), which is suitable for the analysis of an unevenly time coveraged data set.

\subsection{Data selection}
We used three detector data of ASM in order to increase the number of data points.
Since we searched for a periodicity longer than 1 day, we did not carry out the baricentric correction.

We selected data points with a significance: the ratio of the rate to error, $S$ = rate/error, with levels of $\ge$ 0,1,2,3,4 in order to control the data quality.
As well as raw data (no binned data), we also binned a 90 sec integrated data in 1 day, 5 day, 10 day intervals to solve poor statistics. 
In binning, we used a weighted mean and obtained errors by summing in quadrature. 
The power in the Lomb method is linear in the number of data points if a periodicity is real.
As a result, by binning or higher $S$ selection, we obtained a fewer number of data points and  a lower significance for a periodicity. Therefore, we checked for the periodicity under all conditions mentioned above.
Even after binning, some data points are still statistically poor because the original data point distribution is not uniform in time.
Therefore, we selected data points consisting of  $\ge$ 20 raw data points.
We searched for periodicity with a data set of 1700 days.
For Mkn 501, we also search periodicity with a data time limited to MJD 50300-50900, which corresponds to time duration of a long TeV flare.  

\subsection{Result}

We calculated a power spectra with a number of frequencies of 100$\times N$/2 ($N$:number of data).
We limited the search interval for periodicity between 1 day and $T_{obs}$/10. Here, $T_{obs}$ is the total observation time.
We calculated a chance probability with Prob($\ge z$) $\equiv$ $ 1 - (1 - e^{-z} )^N$. Here $z$ is a power.
We detected possible periodicities with a chance probability less than 10$^{-3}$ for Mkn 501, Mkn 421, and PKS 2155-304.

\subsection{Consideration of a gradual increase}

When we applied the fourier transformation or more sophisticated Lomb method to the limited time interval data, we always saw spurious large powers around the test period equal to the observation time. This is due to the low frequency fourier component of Xray data. We have confirmed this effect by simulation. For example, we have generated a linearly growing lightcurve in the observation time, rate=$a$*$t+b$ with error sampled from real data, then we analyzed this data with the Lomb method and found a similar spurious periodicity around $P \sim T_{obs}$. Therefore, to exclude such effects, we removed the low frequency component as follows.
We fitted the lightcurve with a polynomial function with order n=1,2,3,4,5, then we employed the best model, which gave a minimum value of the $\chi^2$/d.o.f, as a global lightcurve. The refined lightcurve obtained by subtracting the modeled function (global lightcurve) from the original lightcurve was analyzed. Furthermore, we limited the periodicity search range between 1 day and $T_{obs}$/10 to reject any spurious periods.
As seen in figure 1, the lightcurve of Mkn 421 shows double peaks with separation of 900 days. Even after subtracting model function, we could not remove the interference between these two peaks and we obtained very large power at 900 days period. Therefore we only used the Mkn 421 data between MJD 50088 (beginning of observation) and MJD 51500 (just before the second peak).

\begin{figure*}[htb]
%\vspace{-60pt}
\psbox[height=6cm,width=7cm]{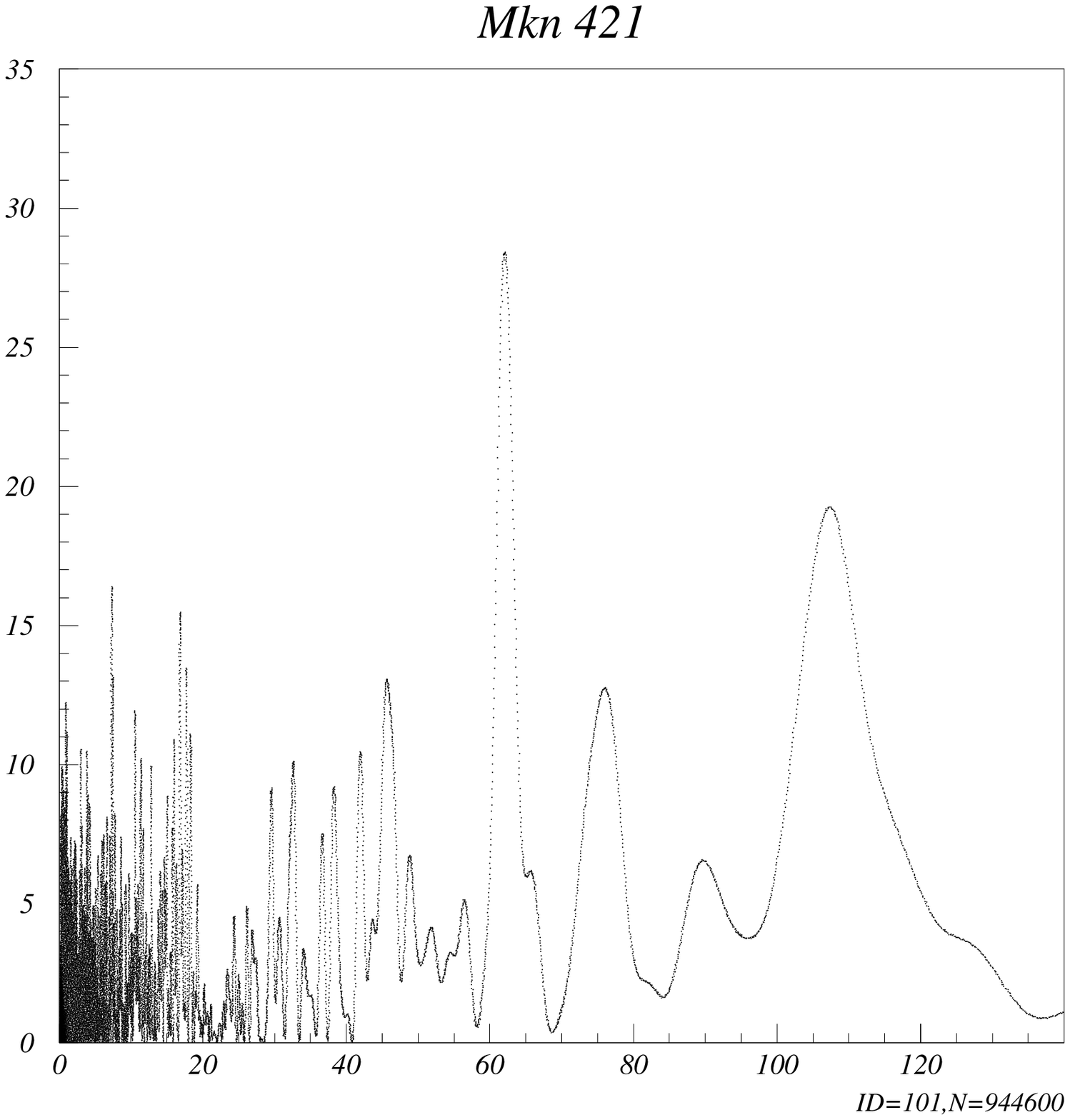}
%\hspace{-35pt}
\psbox[height=6cm,width=7cm]{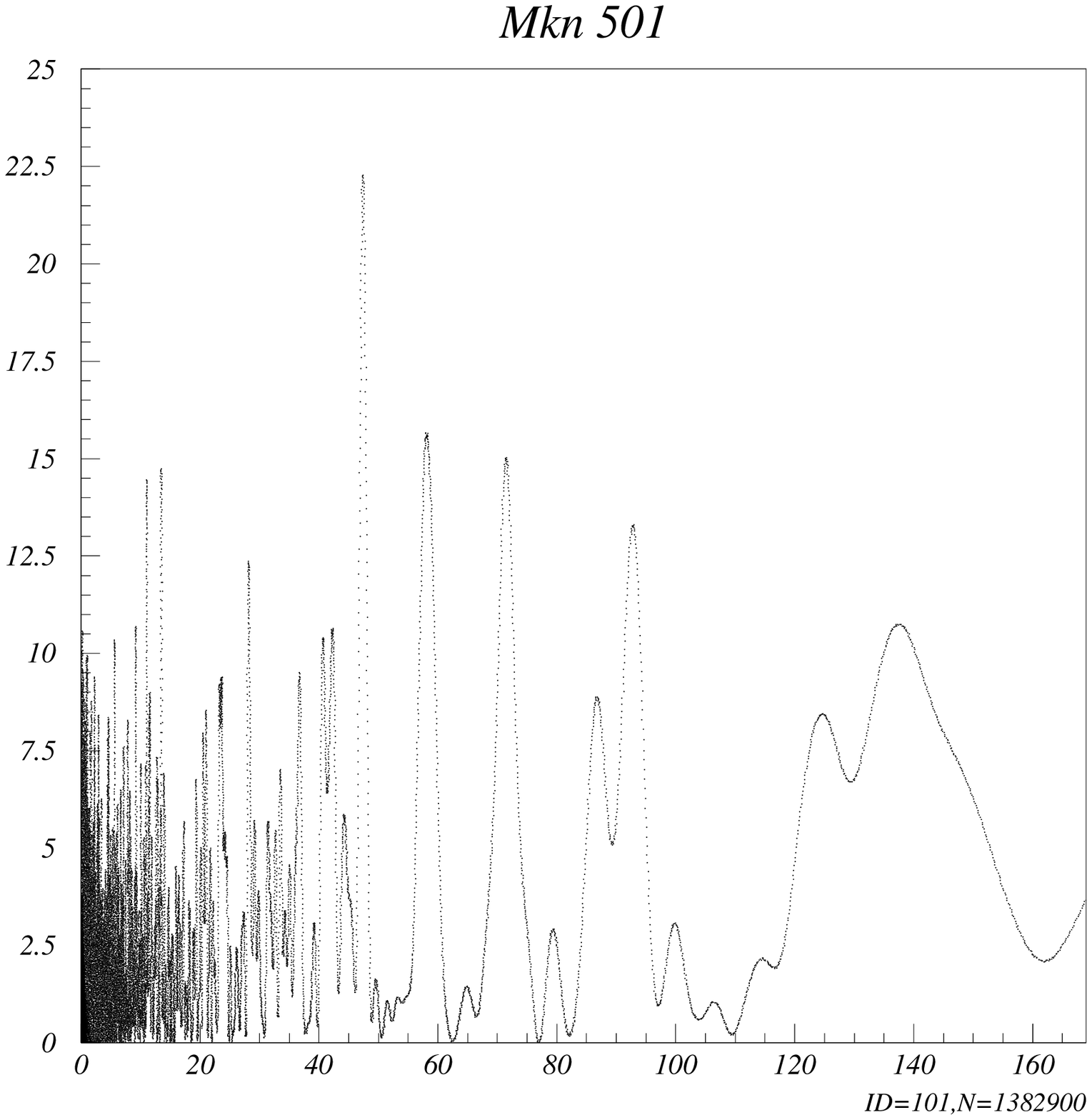}

%\vspace{-60pt}
\psbox[height=6cm,width=7cm]{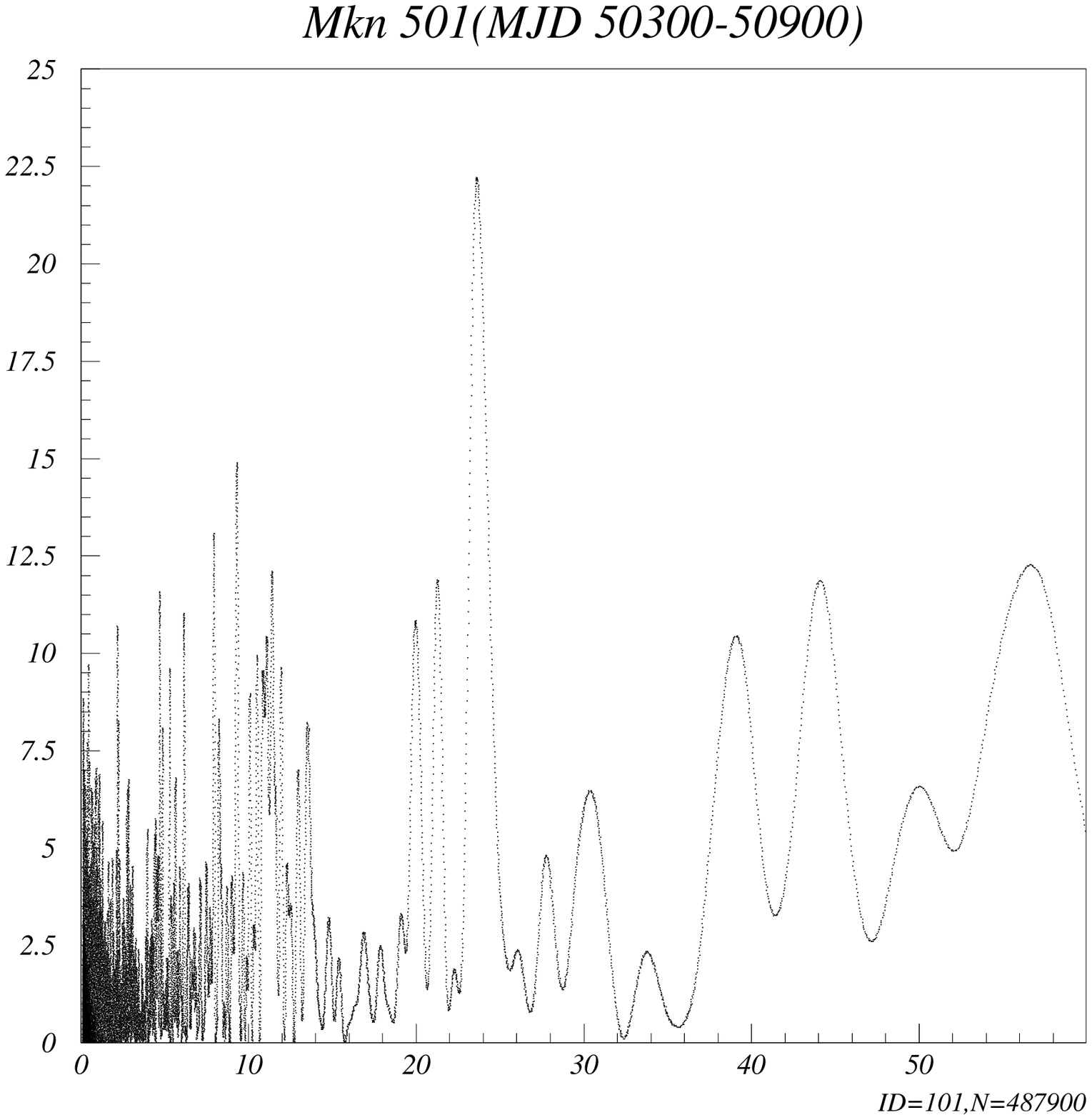}
%\hspace{-35pt}
\psbox[height=6cm,width=7cm]{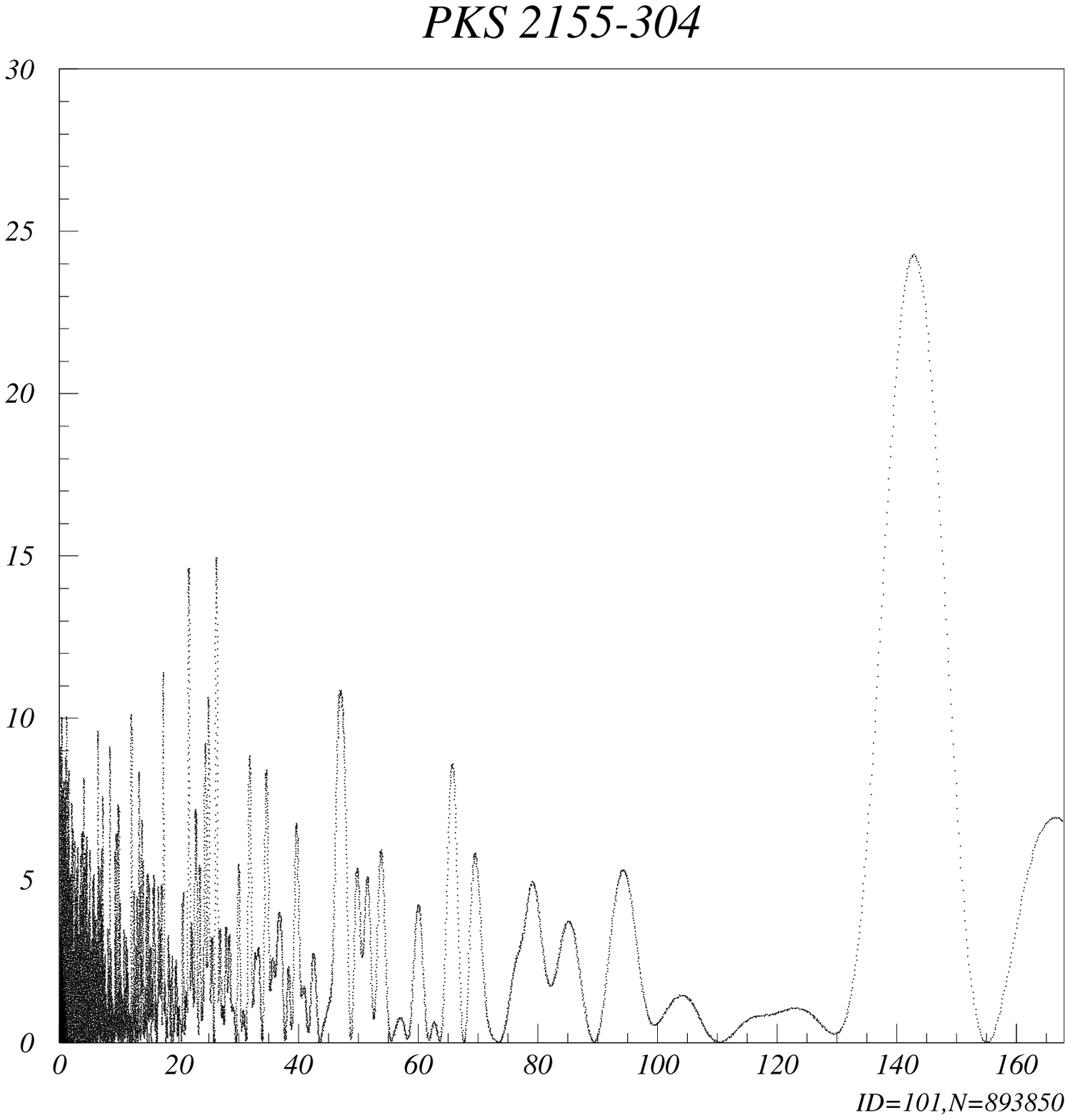}

\caption{Power spectra for Mkn 421, Mkn 501, Mkn 501 during MJD 50300-50900 (around the TeV flare time), PKS 2155-304. Low frequency components have been removed. This data has been obtained from no binning and $S$ $\ge$ 0. Period less than $T_{obs}$/10 are shown. Vertical axis is power on a linear scale, horizon axis is periodicity in unit of day on a linear scale.}
\end{figure*}

\begin{table*}
\caption{Periodicity for all selections. The periodicities which have more than 10 cycles and a chance probability less than 10$^{-3}$ is shown. Low frequency component has been removed.}

\vspace{10pt}
\begin{tabular}{cccrrr} \hline\hline
Target &  binning(day) & $S$ & Period(day) & Chance probability & Number of data  \\ \hline
Mkn 421 & no & 0 & 62.1 & 8.7$\times10^{-9}$ & 18912 \\ 
        &  &  &  107.4 & 8.2$\times10^{-5}$ &    \\ \hline
Mkn 501 & no & 0 & 47.4 & 5.8$\times10^{-6}$ & 27678 \\
Mkn 501(MJD50300-50900)& no  & 0  &  23.6   &  2.2$\times10^{-6}$ & 9778\\
 & no & 1 & 23.6 & 3.9$\times10^{-4}$ & 5157  \\ \hline
PKS 2155-304 & no & 0 & 142.9 & 5.1$\times10^{-7}$ & 17897  \\ 
            & no & 1 & 143.3 & 4.4$\times10^{-6}$ & 7166  \\ \hline
 \end{tabular}
\end{table*}

We show the results in table 2 and figure 2. The many periodicities which have probability with more than 10$^{-3}$ in figure 3 are considered to be statistical fluctuations. 

\section{Discussion}

For Blasers, there is no origin for a long periodicity in a Blackhole system.
However, if multiple Blackholes exist in the center of AGN, we can see a precession or an orbital motion by a secondary Blackhole crossing an accretion disks of the primary or by orbital $\phi$ component.
Hence, we consider three Blasers with significant periodicities as candidate for multiple Blackhole systems.
We discuss parameters of multiple Blackhole systems with the information of the obtained periodicities and discuss a relation between a periodicity in Xrays and TeV gamma rays.

\subsection{Geometry with orbital period}
We consider a two Blackhole system for simplicity.
Assuming a detected periodicity as an orbital period, we estimate a separation distance of two Blackholes as shown in table 3 using the formula for Kepler motion,

\vspace{10pt}
 G$\frac{M_1 + M_2}{r^3}$ = $\bigl( \frac{2\pi}{P} \bigl)^2$.

\vspace{10pt}
\parindent=0pt Here, $M_{\rm 1},M_{\rm 2}$ is each Blackholes mass, $P$ is the orbital period, $r$ is the separation radius.
Even with a current high resolution instrument such as {\it VSOP} ( angular resolution 0.1$\times$10$^{-3}$'') or a next generation {\it VERA} ( 10$\times$10$^{-6}$''), it is not possible to resolve such a two Blackhole system.
We hope for a confirmation of multiple blackholes by a high resolution instrument sometime in the future.

\begin{table}
\caption{Calculated geometry, assuming the detected periodicities caused by an orbital period. $M_{\rm 1}+M_{\rm 2}$ = 2$\times$10$^8M_\odot$ is assumed.($H_0$ = 65 km/(s$\cdot$Mpc)). }

\vspace{10pt}
\begin{tabular}{crrr} \hline\hline
Target & period & separation & angular \\ 
       & (day)  & radius (pc) & distance \\ \hline
Mkn 421 & 62.1    &  9.0$\times10^{-4}$ & 1.3$\times10^{-6}$'' \\
Mkn 501 & 23.6 &  4.6$\times10^{-4}$ & 6.0$\times10^{-7}$'' \\
PKS 2155-304 & 143.0 & 1.5$\times10^{-3}$ & 6.1$\times10^{-7}$'' \\ \hline
 \end{tabular}
\end{table}

\subsection{Constraint on mass ratio}
A precession period for a differentially rotating fluid disk is given by
Papaloizou and Terquem (1995) and the precession period $P_{\rm prec}$ is related to the orbital period $P$ by Larwood (1998) as,

\vspace{10pt}

$\frac{P}{P_{\rm prec}}$ = $\frac{3}{7} \mu \bigl( \frac{1}{1+\mu} \bigl)^{1/2} \bigl( \frac{R_{\rm disk}}{r} \bigl)^{3/2} cos\delta$.

\vspace{10pt}
\parindent=0pt Here, $P_{\rm prec}$ is the primary precession period and $\mu$ is $M_{\rm s}/M_{\rm p}$, $M_{\rm p}$ is the primary blackhole mass, $M_{\rm s}$ is the secondary one, $r$ is the separation distance between the two Blackholes, $R_{\rm disk}$ is the radius of the accretion disk, and $\delta$ is the orbital inclination angle with respect to the disc.

Now, we limit the mass ratio between two blackholes from the detected periodicities, using the above formula. We assume the observed periodicities as  precession periods $P_{prec}$. 
The two Blackhole system with a short orbital period can not be alive for a long time because of  gravitational wave emission. 
When we assume a typical Blackhole separation radius $r/R_{\rm disk}$ $\sim$100, the condition of $\mu\sim10^5$ gives an acceptable orbital period of 0.14 cos$\delta P_{prec}\sim$ 20 days for $P_{prec}$=140 days. If we assume the condition with smaller $\mu$, for example $\mu$=1, we obtained unacceptable orbital period of  3$\times10^{-4}$ cos$\delta P_{prec}$.
We conclude the mass ratio should be very large, typically $\mu \ge10^5$.

On the otherhand, if we assume observed periodicties as orbital periods,
there are two possible models.
One is that we can see the orbital periods by a secondary crossing the accretion disk of the primary or two accretion disk causing a tidal disturbance. This needs a large mass difference for two Blackholes (Sillanpaa et al. 1988;Letho \& Valtonen 1996).
The other is that we can see the orbital period by the orbital $\phi$ motion, which needs two orders mass difference for 23 day periodicity of Mkn 501(Rieger \& Mannheim 2000).

\subsection{Relation with TeV}
It is very interesting that the three Blasers with siginificant periodicities are all TeV gamma sources (Mkn 421;eg.Aharonian et al. 1999b, Mkn 501;eg. Aharonian et al. 1999a, PKS 2155-304;P.M.Chadwick et al. 1999).
If these periodicities are due to a geometrical effect, we expect a similar periodicity in TeV gamma ray intensities.
We searched for such a periodicity using a Lomb method with HEGRA TeV published data, for Mkn 421 during 1997-1998 and for Mkn 501 during 1997-1999 (Lorenz 1999; Aharonian 1999b; Aharonoan 2001;Kranich 2001). We confirmed a 23 day periodicity for Mkn 501 during the 1997 flare time found by Kranich et al.(1999). However, during 1998-1999, 23 day periodicity was not found for Mkn 501. A 62 day periodicty was not found for Mkn 421.
Probably, the sensitivity of TeV gamma ray Cherenkov detectors may not be enough to detect periodicities except around the large flare of Mkn 501 in 1997.

The good coincidence between periodic Blasers and TeV gamma ray Blasers is remarkable.
We can estimate the chance probability of this coincidence at less than 10$^{-2}$.
 There may exist a relation between multiple Blackholes and an electron acceleration to ultra high energy.

\vspace{10pt}
 
{\bf acknowledgements}

We wish to acknowledge the {\it RXTE} ASM group for their public data service and XTEhelp for a kindness in support.
We also acknowledge the HEGRA group, especially Dr. D. Kranich and Dr. H. Krawczynski for giving us a published data set.

\end{document}